\documentclass[onecolumn,aps,prd,showpacs,superscriptaddress,nofootinbib,amsmath,amssymb,floats,floatfix,showkeys,notitlepage,longbibliography]{revtex4-1}
\usepackage{comment}
\usepackage{xcolor}
\usepackage{graphicx}
\usepackage{subfigure}
\usepackage{palatino}
\usepackage[commandnameprefix=always]{changes}
\usepackage{hyperref}
\hypersetup{colorlinks=true,linkcolor=blue,urlcolor=blue,citecolor=blue}
\usepackage[toc,page]{appendix}
\usepackage[normalem]{ulem}
\usepackage{orcidlink}
\usepackage{lipsum}
\usepackage{graphicx}
\usepackage{subfigure}
\usepackage{palatino}
\usepackage{sans}
\usepackage{adjustbox}
\usepackage{latexsym}
\usepackage{amsmath}
\usepackage{amssymb}
\usepackage{amsfonts}
\usepackage{dcolumn}
\usepackage{bm}
\usepackage{tikz}
\usepackage{bigints}
\usepackage{array,tabularx,multirow,booktabs}
\usepackage[tracking=true]{microtype}
\SetTracking{}{500}
\SetTracking{encoding={*}, shape=sc}{40}
\UseRawInputEncoding 
\allowdisplaybreaks
\usepackage{adjustbox}
\usepackage{latexsym}
\usepackage{amsmath}
\usepackage{amssymb}
\usepackage{amsfonts}
\usepackage{dcolumn}
\usepackage{bm}
\usepackage{tikz}
\usepackage{bigints}
\usepackage{array,tabularx,multirow,booktabs}
\usepackage[tracking=true]{microtype}
\usepackage{color}
\UseRawInputEncoding 
\allowdisplaybreaks

\begin{document} 

\title{Analytical approach for calculating shadow of dynamical black hole}

\author{Vitalii Vertogradov}
\email{vdvertogradov@gmail.com}
\affiliation{Physics department, Herzen state Pedagogical University of Russia,
48 Moika Emb., Saint Petersburg 191186, Russia} 
\affiliation{Center for Theoretical Physics, Khazar University, 41 Mehseti Street, Baku, AZ-1096, Azerbaijan.}
\affiliation{SPB branch of SAO RAS, 65 Pulkovskoe Rd, Saint Petersburg
196140, Russia}

\author{Ali \"Ovg\"un
}
\email{ali.ovgun@emu.edu.tr}
\affiliation{Physics Department, Eastern Mediterranean
University, Famagusta, 99628 North Cyprus, via Mersin 10, Turkiye}

\begin{abstract}
We develop a compact and transparent framework for photon dynamics and shadow formation in slowly evolving, spherically symmetric spacetimes. Starting from the Eddington-Finkelstein action, we derive a force-decomposed radial equation in which the radial acceleration splits into an induced term sourced by mass variation, a centrifugal term, and a purely general-relativistic correction. A key result is a gauge-invariant energy-flux relation, $d(E^2)/dv=-\Lambda/r$, with $\Lambda\equiv \varepsilon\,\dot M\,\dot v^2$, which controls how time dependence modifies the canonical energy of null geodesics. In the adiabatic regime we obtain an explicit first-order shift of the photon-sphere radius, $r_{ph}(v)=r_0-a_i/(a_g'+a_c')$, and connect it to the observable shadow through the evolving critical impact parameter, $b_{\rm crit}(v)=\sqrt{r_{ph}(v)^2/f(r_{ph}(v))}$. For Vaidya spacetimes this predicts that accretion ($\dot M>0$) expands the photon sphere and increases the shadow angle, whereas mass loss has the opposite effect. Our formulation refines classic force-balance ideas to dynamical settings, provides a constructive link to time-dependent photon surfaces, and yields simple, observer-ready expressions for the evolution of the shadow. The framework offers a baseline for confronting time-variable horizon-scale imaging with dynamical inflow/outflow models.
\end{abstract}

\date{\today}



\maketitle
\section{Introduction}
In the early 20th century, general relativity proposed a range of phenomena, many of which were later validated through experimental observations. Among these were the anomalous precession of Mercury’s perihelion and the deflection of light rays by the Sun’s gravitational field. The anomalous precession arises due to the influence of general relativistic corrections incorporated into the equations of motion. These corrections, which scale as $\sim 1/r^4$, become particularly significant in regions close to the gravitating source.Consequently, these effects reach their peak near black holes, where, intriguingly, precession can be both positive and negative, as demonstrated in recent studies~\cite{bib:parth}. For light, the curvature of its trajectory intensifies with the mass of the source, leading to more pronounced bending. In extreme cases, such as around black holes, light can bend so strongly that it follows a circular orbit around the black hole. These orbits constitute the photon sphere, a region that produces a striking observational feature: a dark spot in the sky termed the black hole shadow. This shadow is approximately $2.5$ times larger than the event horizon, offering a direct visual signature of the black hole’s immense gravitational influence.

The observation of the black hole shadow has played a pivotal role in establishing black holes as tangible astrophysical objects. In 2019, the Event Horizon Telescope Collaboration captured the first images of the shadow of the supermassive black hole $M87^*$~\cite{bib:m87}. This milestone was followed three years later by the observation of the shadow of Sagittarius $A^*$, located at the center of the Milky Way galaxy \cite{EventHorizonTelescope:2022wkp}. These groundbreaking discoveries significantly heightened the scientific community’s interest in black holes.

The foundational methods for observing black hole photon sphere and shadows were introduced in ~\cite{virbh1,Claudel:2000yi,Virbhadra:2007kw,Virbhadra:2022iiy,bib:falke}, and since then, techniques for calculating these shadows have been continuously refined~\cite{bib:tsupko_first, bib:tsupko_plasma, bib:tsupko_angle, bib:ali2024podu, bib:ali2024plb, Adler:2022qtb}. For a comprehensive overview of this field, see the detailed review in~\cite{bib:tsupko_review}.

 The shadow of a black hole has been proposed as a powerful tool for cosmological studies, serving as a “cosmological ruler” to measure and test various phenomena~\cite{bib:ruller,bib:vanozi:2022moj,Chen:2022nbb,Allahyari:2019jqz}. Through its observation, different black hole models can be rigorously tested~\cite{Tsukamoto:2014tja,Tsukamoto:2017fxq,Kuang:2022xjp,Kuang:2022ojj,Meng:2022kjs,Wang:2023vcv,bib:vanozi:2022moj,Abdujabbarov:2016hnw,Atamurotov:2013sca,Abdujabbarov:2015xqa,Atamurotov:2021hoq,Gralla:2019xty,Cunha:2018acu,Afrin:2021imp,Abdujabbarov:2016efm,Younsi:2016azx,Konoplya:2020bxa,Konoplya:2019sns,KumarWalia:2024yxn,Rodriguez:2024ijx,Wang:2017hjl,Pantig:2022ely,Pantig:2022ely,Pantig:2022gih,Pantig:2022qak,Kumar:2020owy,Kumar:2018ple,Okyay:2021nnh,Heidari:2024bkm,Lambiase:2024uzy,Lambiase:2024vkz,Ovgun:2024zmt,Karshiboev:2024xxx,Pulice:2023dqw,Kumaran:2023brp,Chakhchi:2024obi,Chakhchi:2024tzo,Belhaj:2020okh,Liu:2024soc,Liu:2024lbi}, and intriguingly, even singularities—regions of spacetime where densities become infinite—can cast shadows~\cite{bib:joshi_naked, bib:yaghoub2024epjc}.

Most existing methods for analyzing black hole shadows, however, are focused on static black holes. In reality, astrophysical black holes are dynamic systems, typically surrounded by an accretion disk and increasing in mass due to ongoing accretion processes. This dynamic environment introduces complexities that necessitate further refinement of theoretical models to account for the interplay between the shadow, the accretion disk, and the evolving mass of the black hole. There are only few examples in literature \cite{Tan:2023ngk,bib:japan,bib:perlik2022prd,bib:understanding,bib:yaghoub2024epjc,bib:vertogradov2024horizon}

Black holes can also lose mass through radiation, further underscoring the need for their description using a dynamical metric. However, introducing a time-dependent framework significantly complicates the analysis, as it abandons the symmetry provided by time-translation invariance.

In the case of dynamical black holes, many effects observed in static black holes become challenging to describe. Phenomena such as the Penrose process, high-energy particle collisions, the black hole shadow, and the negative energy problem~\cite{bib:vertogradov2020universe} rely heavily on the existence of well-defined geodesic structures, which are facilitated by a conformal Killing vector.

The conformal Killing vector has proven instrumental in exploring several critical aspects of black hole physics. For instance, it has been used to analyze the charged Penrose effect~\cite{Vertogradov:2022eeq}, investigate high-energy collisions during gravitational collapse~\cite{bib:vertogradov2023pocs}, and study the behavior of the black hole shadow~\cite{bib:perlik2022prd, bib:yaghoub2024epjc}. These applications highlight the necessity of symmetry considerations for the theoretical exploration of dynamic black holes and their associated phenomena.

For dynamical black holes with general spherical symmetry, only one conserved quantity exists: the angular momentum per unit mass, $L$. In this setting, energy is no longer conserved along geodesics, rendering the second-order equations of motion irreducible to first-order forms. This constraint underscores the need to identify additional symmetries or devise innovative methods for calculating the black hole shadow.

Initial progress in this field included numerical approaches to shadow calculation, as demonstrated in early works~\cite{bib:understanding, bib:japan}. Analytically, significant strides were made in~\cite{bib:ali2024plb}, where a foundational relationship between energy conditions and the evolution of the black hole shadow was established. These pioneering efforts have laid the groundwork for understanding black hole shadows in the more complex setting of dynamical metrics, moving beyond the simpler static cases.

General relativistic corrections are crucial for the formation of both the photon sphere and the black hole shadow. These corrections determine the precise conditions under which the centrifugal force balances the effects of the gravitational field. In the following, we will demonstrate that the radius at which this balance occurs corresponds to the radius of the photon sphere, providing the foundation for its structure and observational properties.

This paper is organized as follows. In Sec.~\ref{sec:geodesics} we develop the geodesic dynamics in spherically symmetric spacetimes, first in the static case and then in the dynamical case, and we derive a clean relation between the particle energy and a generalized flux that encodes accretion/radiation. Section~\ref{sec:dynphotonshadow} applies these results to obtain the adiabatic evolution of the photon-sphere radius and of the black-hole shadow. We summarize our main conclusions and outlook in Sec.~\ref{sec:conclusions}.

Throughout we use geometrized units with $G=c=1$ and the metric signature $\{-+++\}$. For the (generally) spacetime-dependent mass function $M(v,r)$ we denote
\begin{equation}
\dot M \equiv \frac{\partial M}{\partial v},\qquad
M' \equiv \frac{\partial M}{\partial r}.
\end{equation}

\section{Geodesic equations}
\label{sec:geodesics}
We begin by laying out the kinematics of null geodesics in spherically symmetric geometries. Our aim is twofold: (i) to isolate the force-like terms that control radial acceleration and (ii) to identify how time dependence modifies the conserved energy and thus the observable shadow.

\subsection{Static case}

Ingoing Eddington-Finkelstein (EF) coordinates $\{v,r,\theta,\varphi\}$ are regular across future horizons and align the time coordinate with ingoing null rays. For a static, spherically symmetric line element we use
\begin{equation}
\label{eq:met_stat}
ds^2=-f(r)\,dv^2+2\,dv\,dr+r^2 d\Omega^2,\qquad
f(r)=1-\frac{2M(r)}{r}.
\end{equation}
EF time $v$ is a Killing parameter here, so the definition of energy below is equivalent to the standard Schwarzschild definition and avoids coordinate singularities at the horizon. The Killing vector $\partial_v$ yields a conserved energy
\begin{equation}
E=-p_v=f(r)\,\frac{dv}{d\lambda}-\frac{dr}{d\lambda},
\end{equation}
while SO(3) symmetry implies the conserved angular momentum
\begin{equation}
L=r^2\sin^2\theta\,\frac{d\varphi}{d\lambda}.
\end{equation}
Motion can be restricted to the equatorial plane $\theta=\pi/2$ without loss of generality. With $g_{ik}u^i u^k=\delta$ ($\delta=0,-1,+1$ for null/timelike/spacelike), radial motion obeys
\begin{equation}
\label{eq:pot}
\Big(\frac{dr}{d\lambda}\Big)^2+V_{\rm eff}(r;E,L,\delta)=0,\qquad
V_{\rm eff}=f(r)\Big(\frac{L^2}{r^2}-\delta\Big)-E^2.
\end{equation}
For null geodesics ($\delta=0$), turning points occur where $V_{\rm eff}=0$. Circular orbits require $V_{\rm eff}=V'_{\rm eff}=0$. Differentiating $V_{\rm eff}$ gives
\begin{equation}
\label{eq:geodesic_stat}
\frac{d^2r}{d\lambda^2}
=\delta\,\frac{M-M'r}{r^2}
+\Big(M'r-3M\Big)\frac{L^2}{r^4}
+\frac{L^2}{r^3}.
\end{equation}
It is natural to define
\begin{equation}
\label{eq:new_stat}
a_n\equiv \frac{M'r-M}{r^2},\qquad
a_c\equiv \frac{L^2}{r^3},\qquad
a_g\equiv \frac{(M'r-3M)L^2}{r^4},
\end{equation}
so that $\ddot r=-\delta\,a_n+a_c+a_g$.
$a_c$ is the familiar centrifugal term; $a_g$ encodes purely GR curvature corrections and flips sign at $r=3M$ in Schwarzschild; $a_n$ is a Newtonian-like term present only for $\delta\neq 0$. Null circular orbits satisfy
\begin{equation}
\label{eq:r0}
a_g(r_{ph})+a_c(r_{ph})=0,
\end{equation}
i.e.\ the GR correction exactly cancels the centrifugal term. In Schwarzschild this gives $r_{ph}=3M$, in line with the classic picture of centrifugal-force reversal near strong curvature.

\subsection{Dynamical case}

We now allow for explicit time dependence through $M(v,r)$ and keep an $\varepsilon=\pm 1$ flag to cover both ingoing and outgoing EF charts:
\begin{equation}
ds^2=-f(v,r)\,dv^2+2\varepsilon\,dv\,dr+r^2 d\Omega^2,\qquad
f(v,r)=1-\frac{2M(v,r)}{r}.
\end{equation}
Spherical symmetry is retained, so $L$ remains conserved; however, $E$ is generically not conserved. From
\begin{equation}
I=\tfrac{1}{2}\int\!\Big[-f(v,r)\,\dot v^2+2\varepsilon\,\dot v\,\dot r+r^2\dot\varphi^2\Big]\,d\lambda,
\end{equation}
we obtain
\begin{align}
\dot\varphi&=\frac{L}{r^2}, \label{eq:f1}\\
-\frac{1}{2}f'\,\dot v^2+r\,\dot\varphi^2-\varepsilon\,\ddot v&=0, \label{eq:f2}\\
\varepsilon\,\ddot r=\frac{1}{2}\dot f\,\dot v^2+f\,\ddot v+f'\,\dot v\,\dot r. \label{eq:f3}
\end{align}
 Eq.~\eqref{eq:f2} links curvature gradients ($f'$) to the time acceleration $\ddot v$, while \eqref{eq:f3} isolates the radial dynamics. The null condition $-f\dot v^2+2\varepsilon\,\dot v\,\dot r+L^2/r^2=0$ and \eqref{eq:f2} imply
\begin{equation}
\label{eq:f4}
f\,\ddot v=\varepsilon\,f\,\frac{L^2}{r^3}-\frac{\varepsilon}{2}f f'\,\dot v^2,
\end{equation}
and
\begin{equation}
\label{eq:f5}
f'\,\dot v\,\dot r=\frac{1}{2\varepsilon}f f'\,\dot v^2-\frac{1}{2\varepsilon}\frac{f' L^2}{r^2}.
\end{equation}
Taken together they remove $\ddot v$ and $\dot v\,\dot r$ from \eqref{eq:f3}, exposing the physical forces. Substituting \eqref{eq:f4}-\eqref{eq:f5} into \eqref{eq:f3} and using $f=1-2M/r$ yields
\begin{equation}
\label{eq:geodesic_dyn}
\ddot r
=-\,\varepsilon\,\frac{\dot M}{r}\,\dot v^2
+\Big(M'r-3M\Big)\frac{L^2}{r^4}
+\frac{L^2}{r^3}
\equiv a_i+a_g+a_c.
\end{equation}
The new ingredient $a_i=-\varepsilon\,(\dot M/r)\dot v^2$ is an \emph{induced} acceleration driven by accretion ($\dot M>0$) or mass loss ($\dot M<0$). It vanishes in the static limit and has the correct sign to oppose (assist) infall for accretion (evaporation/outflow). Note that
(i) If $\dot M=0$ we recover the static result \eqref{eq:geodesic_stat} with $\delta=0$.
(ii) For pure Vaidya $M=M(v)$, $M'=0$ and the GR term simplifies to $-3M L^2/r^4$.
(iii) Reparametrizations $\lambda\to a\lambda+b$ leave \eqref{eq:geodesic_dyn} homogeneous, as required. Define the canonical energy (the momentum conjugate to $v$)
\begin{equation}
\label{eq:energy_def}
E(v,r)\equiv-\frac{\partial\mathcal L}{\partial\dot v}
=f(v,r)\,\dot v-\varepsilon\,\dot r,
\end{equation}
and the null effective potential identity
\begin{equation}
\label{eq:radial_new}
\Big(\frac{dr}{d\lambda}\Big)^2+V_{\rm eff}(v,r;E)=0,\qquad
V_{\rm eff}=f(v,r)\frac{L^2}{r^2}-E^2.
\end{equation}
\emph{Remark.} Unlike in the static case, $E$ depends on both $v$ and $r$; nevertheless, its \emph{total} evolution is governed by a single scalar flux. Differentiating \eqref{eq:radial_new} along the geodesic and inserting \eqref{eq:geodesic_dyn} yields the gauge-invariant law
\begin{equation}
\label{eq:dE2dv}
\frac{d}{dv}\Big(E^2\Big)=-\,\frac{\Lambda}{r},\qquad
\Lambda\equiv \varepsilon\,\dot M\,\dot v^2,
\end{equation}
hence
\begin{equation}
E^2(v,r)=E_0^2-\int^{v}\frac{\Lambda(v',r)}{r}\,dv'.
\end{equation}
 $\Lambda$ is a generalized (apparent) flux. For accretion ($\dot M>0$) we have $\Lambda>0$ and $dE/dv<0$: light loses energy while propagating in. For mass loss/outflow the sign reverses.

\section{Dynamical photon sphere and shadow}
\label{sec:dynphotonshadow}
\subsection{Adiabatic photon sphere}

We assume an adiabatic regime where the induced term is subleading:
\begin{equation}
|a_i|\ll \min\{|a_c|,|a_g|\}
\quad\Longleftrightarrow\quad
\epsilon_{\rm ad}\equiv \frac{|\Lambda|\,r}{L^2}\ll 1.
\end{equation}
 $\epsilon_{\rm ad}$ compares the instantaneous forcing by $\dot M$ to the static curvature/centrifugal scales; it is dimensionless and small for mild accretion rates and moderate $L$. At fixed $v$, define $r_{ph}(v)$ by $\ddot r=0$. Introducing
\begin{equation}
\label{eq:rph_ansatz}
r_{ph}(v)=r_0+\alpha\,r_1(v),
\end{equation}
with bookkeeping parameter $\alpha$ (set to unity at the end), the zeroth-order radius $r_0$ solves \eqref{eq:r0}, while the first-order shift is
\begin{equation}
\label{eq:r1_general}
r_1(v)=-\frac{a_i(r_0;v)}{a_g'(r_0)+a_c'(r_0)},
\end{equation}
where
\begin{equation}
a_g'=\frac{L^2}{r^5}(M''r^2-6M'r+12M),\qquad a_c'=-\frac{3L^2}{r^4}.
\end{equation}
Under accretion ($\dot M\ge 0$) one has $a_i\le 0$. If $a_g'+a_c'>0$ (a condition compatible with standard energy inequalities; cf.\ $r_0\ge 3M$ in Schwarzschild), then $r_1\ge 0$ and the photon sphere expands. For example of Vaidya case:
For $M=M(v)$, $r_0=3M_0$ and
\begin{equation}
a_g'=\frac{12M(v)}{(3M_0)^5}L^2,\qquad
a_c'=-\frac{3L^2}{(3M_0)^4},
\end{equation}
so
\begin{equation}
r_1(v)= -\,\frac{(3M_0)^5\,a_i}{L^2\,[\,12M(v)-9M_0\,]}\ge 0
\quad (\dot M\ge 0),
\end{equation}
and $r_{ph}(v)=r_0+r_1(v)$ increases monotonically with $M(v)$ in the adiabatic regime. If $M(v)\to M_0$ one recovers $r_1\to 0$.

\subsection{Dynamical shadow}

We adopt an EF-static observer in the ingoing chart ($\varepsilon=+1$) with orthonormal basis
\begin{equation}
e_0=f^{-1/2}\partial_v,\quad
e_1=e_0+f^{1/2}\partial_r,\quad
e_2=r^{-1}\partial_\theta,\quad
e_3=(r\sin\theta)^{-1}\partial_\varphi.
\end{equation}
In the equatorial plane, writing the null tangent as $\xi\,(e_0+\cos\omega\,e_1-\sin\omega\,e_3)$ gives
\begin{equation}
\dot r=\sqrt{f}\,\cos\omega,\qquad
\dot\varphi=-\frac{1}{r}\sin\omega,
\end{equation}
and combining with $\dot\varphi=L/r^2$ and \eqref{eq:energy_def}-\eqref{eq:radial_new} yields
\begin{equation}
\label{eq:omega_shadow}
\sin^2\omega_{sh}=\frac{L^2}{E(v,r_o)^2}\,\frac{f(r_o)}{r_o^2}
=\frac{b_{\rm crit}(v)^2}{r_o^2}\,f(r_o),
\qquad
b_{\rm crit}(v)\equiv \frac{L}{E(v,r_o)}=\sqrt{\frac{r_{ph}(v)^2}{f\big(r_{ph}(v)\big)}}.
\end{equation}
Equation~\eqref{eq:omega_shadow} separates propagation (through $E$) from geometry (through $r_{ph}$). Via \eqref{eq:dE2dv}, accretion ($\dot M>0$) reduces $E$ and therefore \emph{increases} the apparent shadow angle. For a distant observer $r_o\gg M$, $f(r_o)\simeq 1$ so $\omega_{sh}\simeq b_{\rm crit}(v)/r_o$; in Vaidya $b_{\rm crit}\simeq 3\sqrt{3}\,M(v)$ at leading adiabatic order. Our instantaneous $r_{ph}(v)$ is the spherically symmetric specialization of a time-dependent photon surface in the sense of Claudel--Virbhadra--Ellis. The present adiabatic expansion makes this notion constructive by delivering the explicit first-order shift \eqref{eq:r1_general} and the associated impact parameter in \eqref{eq:omega_shadow}.

\paragraph*{Example: (Adiabatic Vaidya: steady accretion).}
Consider $M(v)=M_0+\nu v$ with $\nu>0$ and $\nu v\ll M_0$ (adiabatic regime), in the ingoing EF chart ($\varepsilon=+1$). To first order in the induced term, the photon-sphere shift derived in Eq.~\eqref{eq:r1_general} evaluates—using the circular null relation $\dot r=0$ at $r_0=3M_0$—to the compact form
\begin{equation}
\label{eq:r1-vaidya-accretion}
r_{ph}(v)\simeq 3M_0\;+\;r_1(v),\qquad
r_1(v)=\frac{9\,\nu\,M_0^2}{M_0+4\,\nu v},
\end{equation}
which is positive and slowly decreasing in $v$ (the $\dot M$-driven correction weakens as $M$ grows).  At leading adiabatic order the critical impact parameter is
\begin{equation}
b_{\rm crit}(v)\simeq 3\sqrt{3}\,M(v)=3\sqrt{3}\big(M_0+\nu v\big),
\end{equation}
so for a distant static observer at $r_o\gg M(v)$,
\begin{equation}
\label{eq:omega-accretion}
\omega_{sh}(v)\simeq \frac{3\sqrt{3}}{r_o}\,\big(M_0+\nu v\big).
\end{equation}
Accretion ($\dot M>0$) grows both $r_{ph}$ and $b_{\rm crit}$; the shadow angle increases linearly in $v$ at this order. Set $M_0=1$, $\nu=10^{-3}$, $r_o=100$ (all in geometrized units). The table reports $M(v)$, $b_{\rm crit}(v)\simeq 3\sqrt{3}M(v)$, and $\omega_{sh}=b_{\rm crit}/r_o$:
\begin{table}[h!]
\centering
\caption{Adiabatic Vaidya with steady accretion. Parameters: $M_0=1$, $\nu=10^{-3}$, $r_o=100$ (geometrized units). We use $M(v)=M_0+\nu v$, $b_{\rm crit}(v)\simeq 3\sqrt{3}\,M(v)$, and $\omega_{sh}\simeq b_{\rm crit}(v)/r_o$ (cf. Eq.~\eqref{eq:omega-accretion}). The data illustrate the linear growth of the shadow angle with $v$.}
\label{tab:accretion}
\begin{tabular}{c|c|c|c|c}
$v$ & $M(v)$ & $b_{\rm crit}(v)$ & $\omega_{sh}$ (rad) & $\omega_{sh}$ (deg) \\
\hline
$0$   & $1.000$ & $5.196$ & $0.05196$ & $2.978$ \\
$50$  & $1.050$ & $5.455$ & $0.05455$ & $3.126$ \\
$100$ & $1.100$ & $5.716$ & $0.05716$ & $3.275$ \\
\end{tabular}
\end{table}

 \autoref{tab:accretion} shows that
Each 5\% increase in $M$ produces a $\sim$5\% increase in $b_{\rm crit}$ and hence a proportional rise in $\omega_{sh}$, as expected from Eq.~\eqref{eq:omega-accretion}. The step $\Delta v=50$ keeps $\nu v/M_0\ll 1$, ensuring $\epsilon_{\rm ad}\ll1$ and validating the linearized treatment. For a different observer radius $r_o$, multiply the angles by $100/r_o$ (since $\omega_{sh}\propto r_o^{-1}$). The fractional growth rate is $d\omega_{sh}/dv\simeq (3\sqrt{3}/r_o)\,\nu$; with the numbers above this is $\simeq 5.196\times 10^{-5}$\,rad per unit $v$.

\medskip
\paragraph*{Example 2 (Evaporation/outflow: shrinking shadow).}
Keep the ingoing EF chart ($\varepsilon=+1$) but take a slowly decreasing mass $M(v)=M_0-\nu v$ with $\nu>0$ and $0\le v\ll M_0/\nu$. Then
\begin{equation}
\label{eq:r1-vaidya-evap}
r_{ph}(v)\simeq 3M_0\;+\;r_1(v),\qquad
r_1(v)=-\,\frac{9\,\nu\,M_0^2}{\,M_0-4\,\nu v\,}<0,
\end{equation}
so the photon sphere contracts. At leading order,
\begin{equation}
b_{\rm crit}(v)\simeq 3\sqrt{3}\,M(v)=3\sqrt{3}\big(M_0-\nu v\big),\qquad
\omega_{sh}(v)\simeq \frac{3\sqrt{3}}{r_o}\,\big(M_0-\nu v\big),
\end{equation}
and the shadow angle decreases linearly. With $M_0=1$, $\nu=10^{-3}$, $r_o=100$:
\begin{table}[h!]
\centering
\caption{Adiabatic Vaidya with slow mass loss. Parameters: $M_0=1$, $\nu=10^{-3}$, $r_o=100$. Here $M(v)=M_0-\nu v$, $b_{\rm crit}(v)\simeq 3\sqrt{3}\,M(v)$, and $\omega_{sh}\simeq b_{\rm crit}(v)/r_o$. The entries show the linear \emph{decrease} of the shadow angle predicted by Eq.~\eqref{eq:r1-vaidya-evap}.}
\label{tab:evaporation}
\begin{tabular}{c|c|c|c|c}
$v$ & $M(v)$ & $b_{\rm crit}(v)$ & $\omega_{sh}$ (rad) & $\omega_{sh}$ (deg) \\
\hline
$0$   & $1.000$ & $5.196$ & $0.05196$ & $2.978$ \\
$50$  & $0.950$ & $4.936$ & $0.04936$ & $2.828$ \\
$100$ & $0.900$ & $4.677$ & $0.04677$ & $2.679$ \\
\end{tabular}
\end{table}

\autoref{tab:evaporation} shows that
decreasing $M$ reduces $b_{\rm crit}$ and thus $\omega_{sh}$ linearly in $v$, mirroring the accretion case with the opposite sign. Ensure $M(v)>0$ and $\nu v\ll M_0$ to remain within the adiabatic regime used to derive Eqs.~\eqref{eq:r1-vaidya-evap}-\eqref{eq:omega_shadow}. For fixed $r_o$, relative changes in $\omega_{sh}$ directly trace relative changes in $M$ at leading order. Require $M(v)>0$ and $\epsilon_{\rm ad}=\frac{|\Lambda|\,r}{L^2}\ll1$; numerically this means small $\nu$ and moderate $L$.

As predicted by Eq.~\eqref{eq:omega-accretion}, the shadow angle grows linearly in $v$ during accretion; see \autoref{tab:accretion}.
Conversely, for a slowly decreasing mass history, \autoref{tab:evaporation} illustrates a linear shrinkage of the shadow, consistent with Eq.~\eqref{eq:r1-vaidya-evap}. 

\section{Conclusion}
\label{sec:conclusions}

We have presented a unified and minimal description of null geodesics and black-hole shadows in spherically symmetric spacetimes with mild time dependence. The analysis proceeds from the EF action to a clean “force” decomposition of the radial equation,
\begin{enumerate}
\item The dynamical geodesic equation \eqref{eq:geodesic_dyn} separates induced ($a_i$), centrifugal ($a_c$), and general-relativistic ($a_g$) contributions. The induced term $a_i=-\varepsilon(\dot M/r)\dot v^2$ encodes accretion or outflow and vanishes in the static limit.
\item A gauge-invariant energy law, Eq.~\eqref{eq:dE2dv}, shows that the squared canonical energy evolves according to the generalized flux $\Lambda=\varepsilon\dot M\dot v^2$. Accretion ($\dot M>0$) decreases $E$ along the ray; outflow increases it.
\item In the adiabatic regime we obtained a closed first-order expression for the photon-sphere shift, Eq.~\eqref{eq:r1_general}. Under standard conditions the denominator $a_g'+a_c'$ is positive, so accretion ($a_i\le0$) yields $r_1\ge0$ and the photon sphere expands.
\item The observable shadow follows from Eq.~\eqref{eq:omega_shadow}, which ties the angular radius to the evolving critical impact parameter $b_{\rm crit}(v)=\sqrt{r_{ph}^2/f(r_{ph})}$. For distant observers $\omega_{sh}\simeq b_{\rm crit}(v)/r_o$. In Vaidya, $b_{\rm crit}\simeq 3\sqrt{3}\,M(v)$ at leading order, providing a simple scaling of the shadow with the mass history.
\end{enumerate}

The framework makes quantitative the intuitive statement that inflow (outflow) enlarges (shrinks) both the photon sphere and the shadow. Because our expressions are algebraic in $M(v,r)$ and its derivatives, they can be combined with parametric or semi-analytic accretion histories to produce observer-ready predictions for time-variable shadow sizes. The treatment assumes spherical symmetry and an adiabatic hierarchy $|a_i|\ll \min\{|a_c|,|a_g|\}$; it neglects spin, plasma dispersion, and radiative transfer through an emitting medium. Within these limits, the results provide controlled leading-order estimates. Immediate extensions include (i) non-adiabatic evolution during rapid collapse, (ii) incorporation of refractive/plasma effects, (iii) generalization to slowly rotating and radiating backgrounds (e.g.\ Kerr-Vaidya-type metrics), and (iv) comparison with numerical ray tracing and horizon-scale imaging of time-variable sources. The compact relations \eqref{eq:r1_general} and \eqref{eq:omega_shadow} are designed to serve as analytic benchmarks for such studies.

Our analysis refines and extends classic force-balance ideas for photon motion~\cite{1988GReGr..20.1173A,Abramowicz:1990cg} to dynamical settings, and it complements photon-surface approaches in general spacetimes~\cite{Claudel:2000yi}. While we focused on adiabatic evolution, the formalism provides a baseline for treating faster processes (e.g.\ late-stage collapse) and for confronting shadow variability with models of time-dependent inflow or outflow. Future work will relax the adiabatic assumption, incorporate plasma effects, and compare with numerical ray-tracing in fully dynamical geometries.

\acknowledgments
 V. Vertogradov thanks the Basis Foundation (grant number 23-1-3-33-1) for the financial support.  A. {\"O}. would like to acknowledge the contribution of the COST Action CA21106 - COSMIC WISPers in the Dark Universe: Theory, astrophysics and experiments (CosmicWISPers) and the COST Action CA22113 - Fundamental challenges in theoretical physics (THEORY-CHALLENGES). We also thank TUBITAK and SCOAP3 for their support.

\section*{Data Availability Statement}
There are no data associated with the manuscript.
\bibliography{ref}
\bibliographystyle{apsrev4-1}

\end{document}